\documentclass[]{spie}

\usepackage{amsmath,amsfonts,amssymb}
\usepackage{graphicx}
\usepackage[colorlinks=true, allcolors=blue]{hyperref}

\usepackage[numbers]{natbib}
\usepackage{subfigure}
\usepackage[percent]{overpic}
\usepackage{graphicx}
\usepackage{caption}
\usepackage{adjustbox}
\usepackage{amsmath}
\usepackage{upgreek}
\usepackage{lineno}
\makeatletter

\newcommand{\Rmnum}[1]{\expandafter\@slowromancap\romannumeral #1@}
\makeatother
\usepackage{booktabs}

\title{Global Alignment Reference Strategy for Laser Interference Lithography Pattern Arrays}

\author[1]{Xiang Gao}
\author[1, 2]{Jingwen Li}
\author[1]{Zijian Zhong}
\author[*1,2]{Xinghui Li}
\affil[1]{Shenzhen International Graduate School, Tsinghua University,  University Town of Shenzhen, Nanshan District, Shenzhen, 518055, Guangdong, China}
\affil[2]{Tsinghua-Berkeley Shenzhen Institute, Tsinghua University,  University Town of Shenzhen, Nanshan District, Shenzhen, 518055, Guangdong, China}
\affil[*]{Corresponding author: \href{mailto:li.xinghui@sz.tsinghua.edu.cn}{li.xinghui@sz.tsinghua.edu.cn}}

\begin{document} 
\maketitle
\begin{abstract}
Large-area gratings play a crucial role in various engineering fields. However, traditional interference lithography is limited by the size of optical component apertures, making large-area fabrication a challenging task. Here, a method for fabricating laser interference lithography pattern arrays with a global alignment reference strategy is proposed. This approach enables alignment of each area of the laser interference lithography pattern arrays, including phase, period, and tilt angle. Two reference gratings are utilized: one is detached from the substrate, while the other remains fixed to it. To achieve global alignment, the exposure area is adjusted by alternating between moving the beam and the substrate. In our experiment, a 3 $\times$ 3 regions grating array was fabricated, and the $-1$st-order diffraction wavefront measured by the Fizeau interferometer exhibited good continuity. This technique enables effective and efficient alignment with high accuracy across any regions in an interference lithography pattern array on large substrates. 
\end{abstract}

\keywords{Large-area pattern array, Laser interference lithography, Pattern alignment, Reference grating, Large-area grating, Moiré fringes}

\section{INTRODUCTION}
The ability to manipulate, position, and fabricate high-quality structures and materials with nanometer-scale accuracy over large substrate areas is essential for the advancement of nanotechnology in practical applications \cite{cotA,ma2024nanofabrication}. To achieve consistent and uniform surface effects, maintaining the arrangement and spacing of microstructures is crucial \cite{barad2021large}. Microscopic structures dictate macroscopic properties, and controlled periodic structures have applications in various fields, such as organic electronics \cite{ji2020patterning}, biochips, biosensors \cite{biochips}, optical sensing \cite{cheng2018high}, multifunctional film \cite{film}, photonic crystal waveguides \cite{2004spatial}, functional surfaces \cite{wu2019large, shen2021, huerta2017, voisiat2019}.

Laser interference lithography (LIL) facilitates the straightforward, flexible, and rapid production of high-resolution periodic structures across extensive areas without the need for masks. This technique is capable of processing a variety of controllable periodic textured structures, such as periodic nanoparticles, dot arrays, hole arrays, and stripes \cite{liu2023laser}. It can even be applied to curved substrates, taking advantage of its substantial depth of focus \cite{zhou2016}. Additionally, some pulsed, high-power laser interference beams can directly interact with material surfaces through photothermal or photochemical mechanisms, enabling the generation of periodic pattern arrays \cite{marczak2013direct, lasagni2024}. Interference lithography can also be integrated with other fabrication methods, such as soft lithography (SL), chemical etching, and lift-off evaporation (LIFE) technology, to create periodic structures \cite{ji2020patterning}.

This paper presents a method for aligning large-area stripes (grating) arrays. Fabricating these arrays requires multiple exposures across different regions, while ensuring continuous diffraction wavefronts. The fabricated large-area grating arrays can be applied to large spectrometers in astronomical telescopes \cite{Telescope, steidel2022design}, high-power chirped pulse amplification laser systems for inertial confinement fusion \cite{LaserFusion, zuegel2006laser}, and long-range grating interferometers \cite{xing2017spatially, shimizu2021laser}. 

In previous methods, laser interference lithography allows for the fabrication of meter-size holographic gratings in a single exposure \cite{Bonod:16}. However, the size of the exposure aperture is limited by the optical components, as manufacturing large-aperture, low-aberration collimating lenses is both challenging and expensive. Schattenburg et al. have proposed a scanning exposure method \cite{heilmann2001digital, SCAN, chen2002nanometer}. In this method, the substrate moves continuously within the exposure region while the phase of the interference fringes is adjusted by an acousto-optic modulator. This adjustment maintains the interference pattern in a stationary position relative to the substrate. Typically, scanning beam interference lithography systems require complex and precise control techniques. 

In contrast, using multiple exposures to create grating arrays is a more convenient and cost-effective method. To ensure the continuity of the diffraction wavefront of the grating array, it is crucial to ensure that the fringe phase, period, and tilt angle remain consistent across different exposure regions. Turukhano et al. have proposed a method for fabricating gratings array in a single direction using a reference grating to monitor alignment errors \cite{TURUKHANO1996263}. However, this approach is restricted to the production of long-strip gratings and cannot be extended in the perpendicular direction. Shi et al. proposed a method based on the moiré fringes generated by latent gratings (photoresist that is exposed but not developed) to monitor alignment errors between different exposure regions \cite{shi2010fabrication, shi2011fabrication}. However, using latent fringe patterns to monitor errors can lead to more complex optical systems and more challenging optical adjustments. Additionally, using latent gratings as references is constrained by the specific patterns (designs, materials) being processed. In contrast, reference gratings can be applied to a wider range of scenarios.

Since several methods for fabricating unlimited long strip gratings have been reported, such as Turukhano et al.'s mosaic method and Ma et al.'s broad-beam scanning method \cite{TURUKHANO1996263, ma2017achieving}, obtaining unlimited long reference gratings is feasible. In this paper, we propose a method that uses two orthogonally placed strip reference gratings, with the similar period as the grating being fabricated. The moiré fringes generated by the reference gratings are employed to monitor alignment errors. One reference grating is separated from the substrate and the exposure region is adjusted by moving the beam. The other reference grating is fixed relative to the substrate, and changes to the exposure region are achieved by simultaneously moving both the substrate and the reference grating. The high diffraction efficiency of the reference gratings makes it easier to adjust and observe moiré fringes, making this method both simple and stable. This method has been experimentally validated for fabricating large-area grating arrays.

\section{Results}

\subsection{Alignment System setup}
\begin{figure} []
	\begin{center}
		\includegraphics[height=10cm]{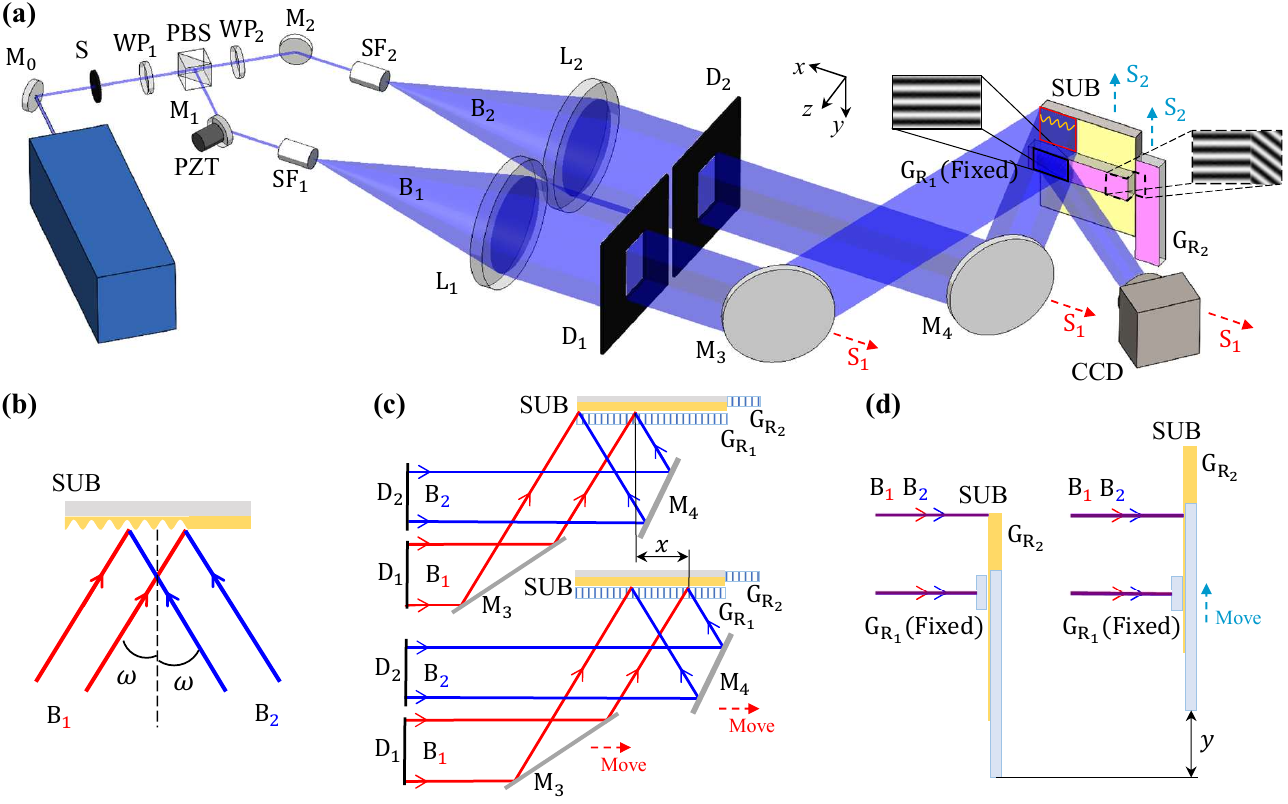}
	\end{center}
	\caption
	{\label{main} Grating pattern array exposure system: (a) System structure; (b) Two-beam interference; (c) Beam movement along the $x$-axis; (d) Substrate movement along the $y$-axis.}
\end{figure} 

Figure \ref{main}(a) illustrates the grating pattern array exposure system we constructed, with the $x$, $y$, and $z$ axes corresponding to the grating vector direction (horizontal), the grating lines direction (vertical), and the grating surface normal direction, respectively. When the shutter S is open, the laser is split into two beams by a polarizing beam splitter (PBS). A half-wave plate $\text{WP}_1$ is used to adjust the intensity ratio between beam $\text{B}_1$ and $\text{B}_2$. After splitting, $\text{B}_2$ is adjusted to match the polarization direction of $\text{B}_1$ using a half-wave plate $\text{WP}_2$, before passing through the spatial filter $\text{SF}_2$. $\text{B}_1$ directly enters the spatial filter $\text{SF}_1$. After beam expansion through the spatial filters, lenses $\text{L}_1$ and $\text{L}_2$ are used for collimation. Then the beams are shaped into rectangular cross-sections using the rectangular apertures $\text{D}_1$ and $\text{D}_2$. Finally, after reflection by mirrors $\text{M}_3$ and $\text{M}_4$, the two exposure beams overlap to form an interference exposure region. If the angle between the beam and the substrate normal is denoted as $\omega$, as shown in Figure \ref{main}(b), the interference between the two beams produces fringes with a period of
\begin{equation}
	d = \frac{\lambda}{2\sin \omega},
\end{equation}
where $d$ represents the period, $\lambda$ is the wavelength of the beams.

The two exposure beams, after being diffracted by the reference grating, will interfere to form moiré fringes with their $-2$nd and $-1$st-order diffracted beams. The resulting moiré fringes are monitored using a CCD camera. A fixed reference grating $\mathrm{G_{R_1}}$ is placed in front of the substrate SUB. During the fabrication process in the $x$ direction, the displacement stage $\text{S}_1$ moves the mirrors $\text{M}_3$, $\text{M}_4$, and the CCD camera simultaneously along the $x$ direction. The movement of mirrors $\text{M}_3$ and $\text{M}_4$ shifts the exposure region, as shown in Figure \ref{main}(c). Another reference grating $\mathrm{G_{R_2}}$ is fixed to the side of the substrate SUB. During the fabrication process in the $y$ direction, the displacement stage $\text{S}_2$ moves both the substrate SUB and the reference grating $\mathrm{G_{R_2}}$ simultaneously along the $y$ direction (Figure \ref{main}(d)). 

The mirror $\text{M}_1$ is connected to a piezoelectric actuator PZT to compensate for phase errors. Mirror $\text{M}_3$ is mounted on a dual-axis piezoelectric stage, used to compensate for period and tilt errors. 

\subsection{Alignment Error Monitoring and Compensation}
\begin{figure} [b]
	\begin{center}
		\includegraphics[height=4cm]{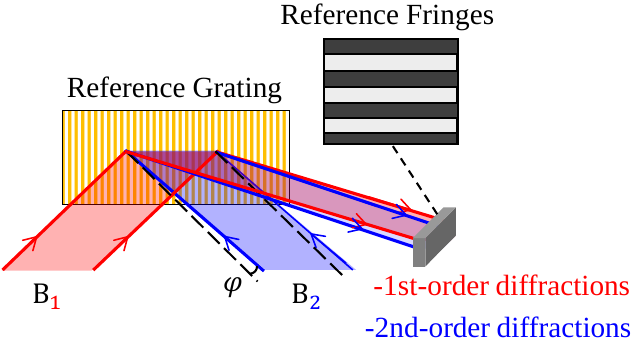}
	\end{center}
	\caption
	{\label{-1-2} Principle of generating reference fringes for alignment errors monitoring}
\end{figure}

During the fabrication process, three main types of errors typically occur: phase errors, period errors, and tilt errors. We use moiré fringes as reference fringes to monitor alignment errors. The principle for generating reference fringes is illustrated in Figure \ref{-1-2}, where the $-2$nd-order diffraction beam of beam $\text{B}_2$ interferes with the $-1$st-order diffraction beam of beam $\text{B}_1$. By introducing a small angle $\varphi$, the reference fringes can be directly observed by the CCD camera.

\begin{figure} [t]
	\begin{center}
		\includegraphics[height=8cm]{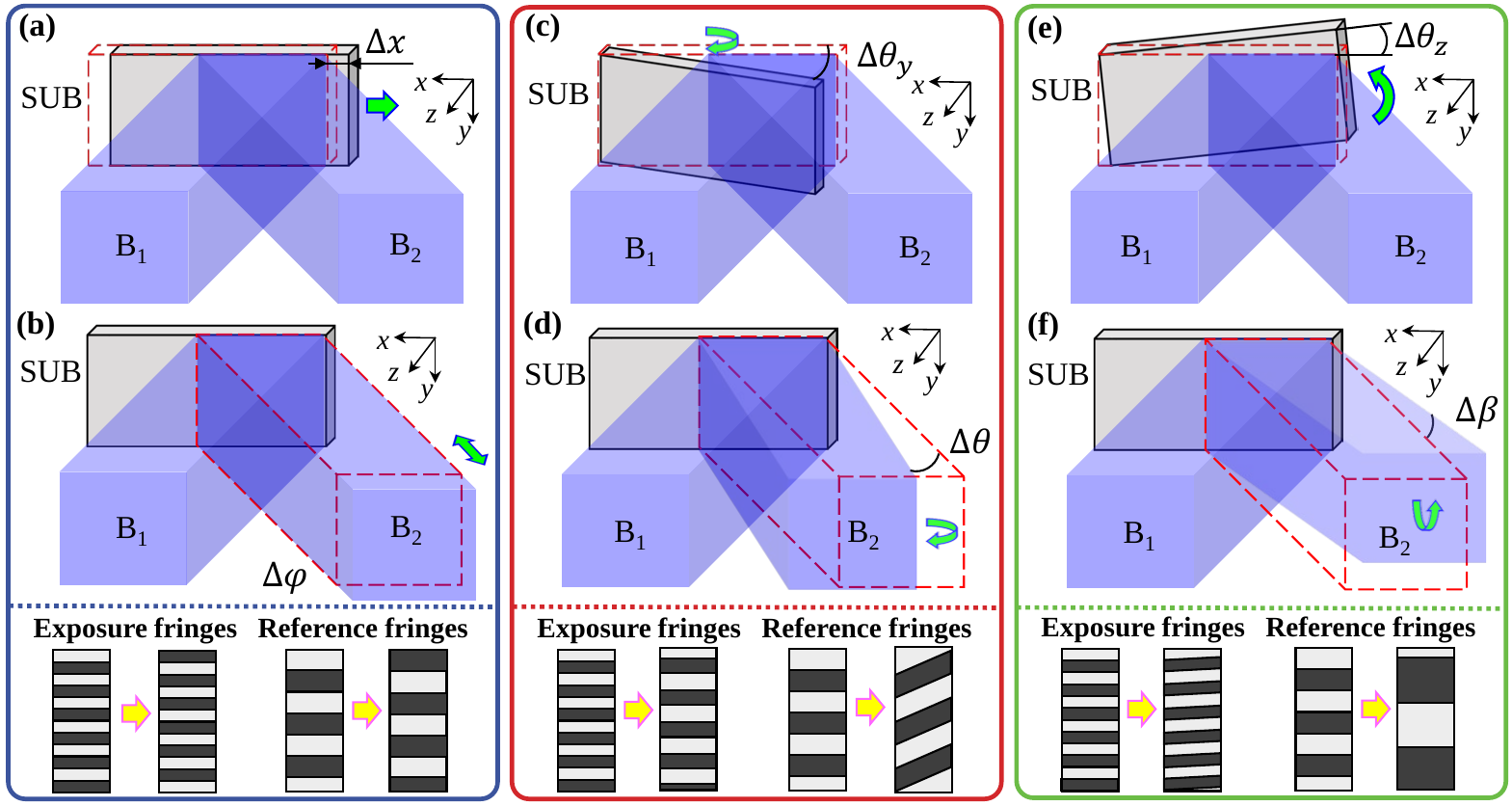}
	\end{center}
	\caption
	{\label{PPT} Alignment errors caused by changes in the position and orientation between the beams and the substrate: (a) Relative displacement between the beams and the substrate along the $x$-axis; (b) Beam phase drift; (c) Relative rotation between the beams and the substrate along the $y$-axis; (d) Change in the angle between the two beams; (e) Relative rotation between the beams and the substrate along the $z$-axis; (f) Deflection between the two beams.}
\end{figure}

The displacement of the substrate relative to the beams along the $x$-axis causes phase errors, as shown in Figure \ref{PPT}(a). The reference fringes will change synchronously with the exposure fringes. That is, the reference fringes move the same period as the exposure fringes. Similarly, phase drift in the exposure beams will also result in phase errors (Figure \ref{PPT}(b)). The phase drift causes the reference fringes and the exposure fringes to change synchronously as well. Therefore, aligning the reference fringes' phase between two exposures ensures the phase continuity of the exposure fringes in different exposure regions.

When a relative deflection occurs between the exposure beams and the substrate along the $y$-axis (Figure \ref{PPT}(c)), it results in period errors in the exposure fringes. Simultaneously, the reference fringes exhibit changes in tilt angle. A slight period error can result in a noticeable tilt angle of the reference fringes. Similarly, when the angle between the two exposure beams changes (Figure \ref{PPT}(d)), it also results in period errors and changes in the tilt angle of the reference fringes. These are the two primary causes of period errors, meaning that changes in the tilt angle of the reference fringes indicate the presence of period errors in the exposure fringes.

Tilt errors also have two primary causes. The substrate's rotation relative to the beam along the $z$-axis (Figure \ref{PPT}(e)) and the relative rotation between two exposure beams (Figure \ref{PPT}(f)). In both cases, the reference fringes will experience changes in period, and a slight tilt error can lead to a noticeable variation in the period of the reference fringes. Therefore, a change in the period of the reference fringes indicates the presence of tilt errors in the exposure fringes.

Based on the above analysis, ensuring consistent phase, period, and tilt angle of the exposure fringes can be achieved by simply maintaining the same phase, period, and tilt angle of the reference fringes as those generated in the previous exposure.

The alignment error is compensated based on real-time monitoring of the reference fringes. Mirror $\text{M}_1$ is mounted on a piezoelectric actuator, which precisely moves the mirror back and forth. This motion alters the optical path of a single beam, enabling fine adjustment of the fringe position in the exposure region to compensate for phase errors. Continuous monitoring of the reference fringes can lock the phase in place. A dual-axis piezoelectric mirror mount is used to control both the yaw and pitch angles of mirror $\text{M}_3$. Adjusting the yaw angle changes the angle between the two exposure beams, which in turn modifies the fringe period, allowing for correction of period errors. Altering the pitch angle of mirror $\text{M}_3$ adjusts the beam's azimuth, thereby compensating for tilt errors. 

\begin{figure} [t]
	\begin{center}
		\includegraphics[height=12cm]{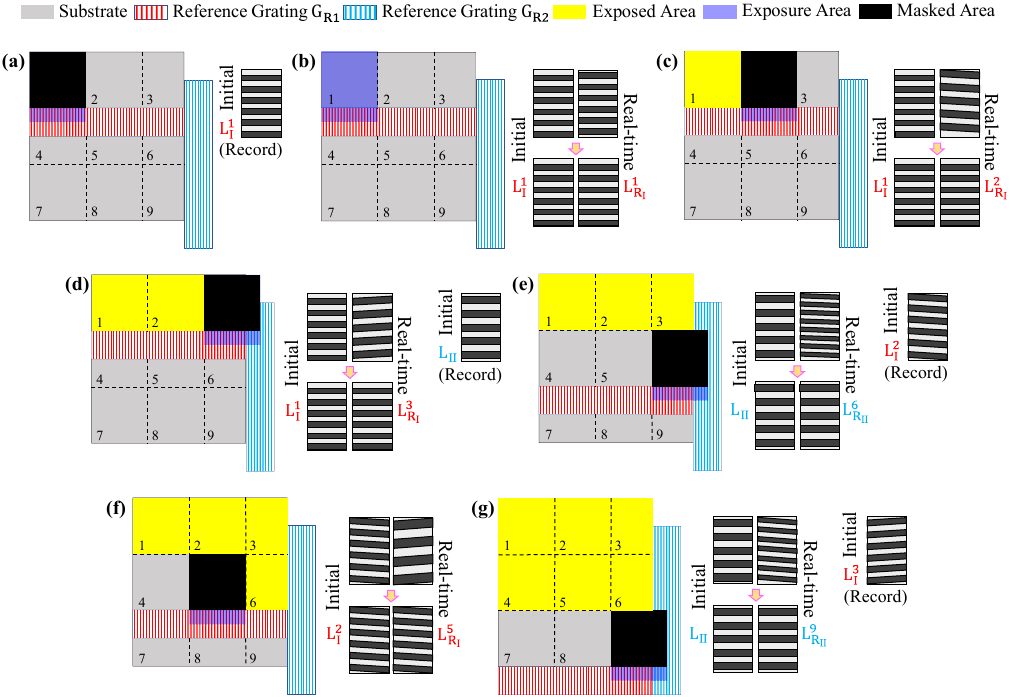}
	\end{center}
	\caption
	{Fabrication procedure for a $3 \times 3$ grating pattern array: (a) Record the initial fringes $\mathrm{L_\Rmnum{1}^1}$; (b) Lock the real-time fringes $\mathrm{L_{R_\Rmnum{1}}^1}$ to the initial fringes $\mathrm{L_\Rmnum{1}^1}$ and expose Region 1; (c) Lock the real-time fringes $\mathrm{L_{R_\Rmnum{1}}^2}$ to the initial fringes $\mathrm{L_\Rmnum{1}^1}$ and expose Region 2; (d) Lock the real-time fringes $\mathrm{L_{R_\Rmnum{1}}^3}$ to the initial fringes $\mathrm{L_\Rmnum{1}^1}$, record the initial fringes $\mathrm{L_{\Rmnum{2}}}$, and expose Region 3; (e) Lock the real-time fringes $\mathrm{L_{R_{\Rmnum{2}}}^6}$ to the initial fringes $\mathrm{L_{\Rmnum{2}}}$, record the initial fringes $\mathrm{L_\Rmnum{1}^2}$, and expose Region 6; (f) Lock the real-time fringes $\mathrm{L_{R_\Rmnum{1}}^5}$ to the initial fringes $\mathrm{L_\Rmnum{1}^2}$ and expose Region 5; (g) Lock the real-time fringes $\mathrm{L_{R_{\Rmnum{2}}}^9}$ to the initial fringes $\mathrm{L_{\Rmnum{2}}}$, record the initial fringes $\mathrm{L_\Rmnum{1}^3}$, and expose Region 9;
	\label{procedure}}
\end{figure} 

We take the example of a 3 $\times$ 3 regions grating pattern array to explain the alignment and exposure procedure (Figure \ref{procedure} and Table \ref{tab}). Before starting the fabrication procedure, the exposure beams need to be aligned with both reference gratings at the same time. By adjusting the pose of the two reference gratings, high-quality reference fringes should be generated from each reference grating. 

Here, we refer to the substrate regions from left to right and top to bottom as Region 1 to Region 9. $\mathrm{L_\Rmnum{1}^\mathit{k}}$ represents the initial fringes recorded by the reference grating $\mathrm{G_{R_1}}$ during the $k$th ($k = 1,2...$) row of the fabrication procedure. $P_\mathrm{\Rmnum{1}}^k$, $\alpha_\mathrm{\Rmnum{1}}^k$, and $D_\mathrm{\Rmnum{1}}^k$ correspond to the phase, tilt angle, and period of the initial fringes $\mathrm{L_\Rmnum{1}^\mathit{k}}$. $\mathrm{L_{\Rmnum{2}}}$ refers to the initial fringes recorded by the reference grating $\mathrm{G_{R_2}}$. $P_\mathrm{\Rmnum{2}}$, $\alpha_\mathrm{\Rmnum{2}}$, and $D_\mathrm{\Rmnum{2}}$ represent the phase, tilt angle, and period of the initial fringes $\mathrm{L_{\Rmnum{2}}}$. $\mathrm{L}_{\text{R}_i}^j$ represents the real-time reference fringes generated by reference grating $i$ ($i = \mathrm{\Rmnum{1}}$ for $\mathrm{G_{R_1}}$; $i = \mathrm{\Rmnum{2}}$ for $\mathrm{G_{R_2}}$) in Region $j$  (j = 1, 2...). $P_{\mathrm{R}_i}^j$, $\alpha_{\mathrm{R}_i}^j$, and $D_{\mathrm{R}_i}^j$ represent the phase, tilt angle, and period of the real-time fringes $\mathrm{L}_{\text{R}_i}^j$.

\begin{table}[htbp]
	\caption{Alignment and exposure procedure for grating pattern array fabrication.}
	\label{tab}
	\begin{tabular*}{\textwidth}{@{}cllccc@{}}
		
		\toprule
		
		Step & Beam position &Action & Control & Record & Expose\\
		\midrule
		1& Region 1 & Record initial fringes $\mathrm{L_\Rmnum{1}^1}$ & - & $\mathrm{L_\Rmnum{1}^1}$ & -\\
		
		& & Expose Region 1 & $\mathrm{L_{R_\Rmnum{1}}^1}$ = $\mathrm{L_\Rmnum{1}^1}$ & -& \checkmark\\
		
		2 & Region 2 (Beam moving)& Expose Region 2 & $\mathrm{L_{R_\Rmnum{1}}^2}$ = $\mathrm{L_\Rmnum{1}^1}$ & - & \checkmark\\
		
		3 & Region 3 (Beam moving)& Record initial fringes $\mathrm{L_{\Rmnum{2}}}$ & $\mathrm{L_{R_\Rmnum{1}}^3}$ = $\mathrm{L_\Rmnum{1}^1}$ & $\mathrm{L_{\Rmnum{2}}}$ & -\\
		
		& & Expose Region 3 & $\mathrm{L_{R_\Rmnum{1}}^3}$ = $\mathrm{L_\Rmnum{1}^1}$ & - & \checkmark\\
		
		4 & Region 6 (Substrate moving)& Record initial fringes $\mathrm{L_\Rmnum{1}^2}$ & $\mathrm{L_{R_{\Rmnum{2}}}^6}$ = $\mathrm{L_{\Rmnum{2}}}$ & $\mathrm{L_\Rmnum{1}^2}$ & -\\
		
		& &Expose Region 6 & $\mathrm{L_{R_{\Rmnum{1}}}^6}$ = $\mathrm{L_\Rmnum{1}^2}$ & - & \checkmark\\
		
		5& Region 5 (Beam moving)& Expose Region 5 & $\mathrm{L_{R_\Rmnum{1}}^5}$ = $\mathrm{L_\Rmnum{1}^2}$ & - & \checkmark \\
		
		6& Region 4 (Beam moving)& Expose Region 4 & $\mathrm{L_{R_\Rmnum{1}}^4}$ = $\mathrm{L_\Rmnum{1}^2}$ & - & \checkmark \\
		
		7& Region 9 (Beam and Substrate moving)& Record initial fringes $\mathrm{L_\Rmnum{1}^3}$ & $\mathrm{L_{R_{\Rmnum{2}}}^9}$ = $\mathrm{L_{\Rmnum{2}}}$ &  $\mathrm{L_\Rmnum{1}^3}$ & -\\
		& & Expose Region 9 & $\mathrm{L_{R_\Rmnum{1}}^9}$ = $\mathrm{L_\Rmnum{1}^3}$ & - & \checkmark\\
		
		8& Region 8 (Beam moving)& Expose Region 8 & $\mathrm{L_{R_\Rmnum{1}}^8}$ = $\mathrm{L_\Rmnum{1}^3}$ & - & \checkmark \\
		
		9& Region 7 (Beam moving)& Expose Region 7 & $\mathrm{L_{R_\Rmnum{1}}^7}$ = $\mathrm{L_\Rmnum{1}^3}$ & - & \checkmark \\
		\bottomrule
	\end{tabular*}
\end{table}

Step 1: Open the shutter S while masking the substrate and record the initial fringes $\mathrm{L_\Rmnum{1}^1}$, as shown in Figure \ref{procedure}(a). Next, unmask the substrate and expose the first region while locking the real-time fringes $\mathrm{L_{R_\Rmnum{1}}^1}$ to the initial fringes $\mathrm{L_\Rmnum{1}^1}$, ensuring that $P_\mathrm{R_\Rmnum{1}}^1 = P_\mathrm{\Rmnum{1}}^1$, $\alpha_\mathrm{R_\Rmnum{1}}^1 = \alpha_\mathrm{\Rmnum{1}}^1$, and $D_\mathrm{R_\Rmnum{1}}^1 = D_\mathrm{\Rmnum{1}}^1$, as shown in Figure \ref{procedure}(b).

Step 2: Move the displacement stage $\text{S}_1$ to align the exposure region with Region 2. Expose Region 2 while locking the real-time fringes $\mathrm{L_{R_\Rmnum{1}}^2}$ to the initial fringes $\mathrm{L_\Rmnum{1}^1}$ (ensuring $P_\mathrm{R_\Rmnum{1}}^2 = P_\mathrm{\Rmnum{1}}^1$, $\alpha_\mathrm{R_\Rmnum{1}}^2$ = $\alpha_\mathrm{\Rmnum{1}}^1$, and $D_\mathrm{R_\Rmnum{1}}^2$ = $D_\mathrm{\Rmnum{1}}^1$), as shown in Figure \ref{procedure}(c).

Step 3: Move the displacement stage $\text{S}_1$ to align the exposure region with Region 3. Here, both reference gratings appear simultaneously in the exposure region. While masking the substrate, open the shutter S and lock the real-time fringes $\mathrm{L_{R_\Rmnum{1}}^3}$ to the initial fringes $\mathrm{L_\Rmnum{1}^1}$ (ensuring $P_\mathrm{R_\Rmnum{1}}^3 = P_\mathrm{\Rmnum{1}}^1$, $\alpha_\mathrm{R_\Rmnum{1}}^3$ = $\alpha_\mathrm{\Rmnum{1}}^1$, and $D_\mathrm{R_\Rmnum{1}}^2$ = $D_\mathrm{\Rmnum{1}}^1$), while simultaneously recording the initial fringes $\mathrm{L_{\Rmnum{2}}}$ (Figure \ref{procedure}(d)). Expose Region 3 with real-time fringes $\mathrm{L_{R_\Rmnum{1}}^3}$ locked to initial fringes $\mathrm{L_\Rmnum{1}^1}$.

Step 4: Move the displacement stage $\text{S}_2$ to align Region 6 of the substrate with the exposure region. The reference grating $\mathrm{G_{R_2}}$ and the substrate move simultaneously. While masking the substrate, open the shutter S and lock the real-time fringes $\mathrm{L_{R_{\Rmnum{2}}}^6}$ generated by the reference grating $\mathrm{G_{R_2}}$ to the initial fringes $\mathrm{L_{\Rmnum{2}}}$ (ensuring $P_\mathrm{R_{\Rmnum{2}}}^6 = P_\mathrm{\Rmnum{2}}$, $\alpha_\mathrm{R_{\Rmnum{2}}}^6 = \alpha_\mathrm{\Rmnum{2}}$, and $D_\mathrm{R_{\Rmnum{2}}}^6 = D_\mathrm{\Rmnum{2}}$), while simultaneously recording the initial fringes $\mathrm{L_\Rmnum{1}^2}$ (Figure \ref{procedure}(e)). Expose Region 6 with real-time fringes $\mathrm{L_{R_\Rmnum{1}}^6}$ locked to initial fringes $\mathrm{L_\Rmnum{1}^2}$.

Step 5: Move the displacement stage $\text{S}_1$ to align the exposure region with Region 5. Expose Region 5 while locking the real-time fringes $\mathrm{L_{R_\Rmnum{1}}^5}$ to the initial fringes $\mathrm{L_\Rmnum{1}^2}$ (ensuring $P_\mathrm{R_\Rmnum{1}}^5 = P_\mathrm{\Rmnum{1}}^2$, $\alpha_\mathrm{R_\Rmnum{1}}^5$ = $\alpha_\mathrm{\Rmnum{1}}^2$, and $D_\mathrm{R_\Rmnum{1}}^5$ = $D_\mathrm{\Rmnum{1}}^2$), as shown in Figure \ref{procedure}(f).

Step 6: Expose Region 4, using a method similar to that of the Step 5, ensuring that the real-time fringes $\mathrm{L_{R_\Rmnum{1}}^4}$ align with the initial fringes $\mathrm{L_\Rmnum{1}^2}$.

Step 7: Move both displacement stages $\text{S}_1$ and $\text{S}_2$ to align the exposure region with Region 9. While masking the substrate, open the shutter S and lock the real-time fringes $\mathrm{L_{R_{\Rmnum{2}}}^9}$ generated by the reference grating $\mathrm{G_{R_2}}$ to the initial fringes $\mathrm{L_{\Rmnum{2}}}$ (ensuring $P_\mathrm{R_{\Rmnum{2}}}^9 = P_\mathrm{\Rmnum{2}}$, $\alpha_\mathrm{R_{\Rmnum{2}}}^9 = \alpha_\mathrm{\Rmnum{2}}$, and $D_\mathrm{R_{\Rmnum{2}}}^9 = D_\mathrm{\Rmnum{2}}$), while simultaneously recording the initial fringes $\mathrm{L_\Rmnum{1}^3}$ (Figure \ref{procedure}(g)). Expose Region 9 with real-time fringes $\mathrm{L_{R_\Rmnum{1}}^9}$ locked to initial fringes $\mathrm{L_\Rmnum{1}^3}$.

Step 8, 9: Expose Region 8 and 7 using the method similar in the Step 5, but with the initial fringes $\mathrm{L_\Rmnum{1}^3}$.

\subsection{Alignment Error Analysis}
A Fizeau interferometer can be used to detect alignment errors. By measuring the $-m$th ($m = 1,2...$) order diffraction wavefront of the grating array, the errors can be analyzed. The 0th-order diffraction wavefront also need to be measured to eliminate the influence of substrate flatness on the diffraction wavefront. When measuring the substrate flatness, the grating is aligned parallel to the reference flat of the Fizeau interferometer. Then rotate the grating to measure the $-m$th order diffraction wavefront, with the angle $\alpha_m$ between the normal of the substrate and the normal of the reference flat given by
\begin{equation}
	\alpha_m = \arcsin\left(\frac{m\lambda_\text{F}}{2d_\text{g}}\right),
	\label{alpha}
\end{equation}
where $m$ represents $-m$th order diffraction wavefront, $\lambda_\text{F}$ is the wavelength of the Fizeau interferometer, and $d_\text{g}$ is the period of the fabricated grating.

The wavefront measured using 0th-order diffracted light is denoted as $w_0(x,y)$, and the wavefront measured using $-m$th order diffracted light is denoted as $w_m(x,y)$. The wavefront after eliminating the influence of the substrate is given by
\begin{equation}
	w = w_m(x,y) - \cos\alpha_m w_0(x\cos\alpha_m,y),
\end{equation}
where $\alpha_m$ is the angle between the normal of the substrate and the normal of the reference flat, as given in Equation \ref{alpha}. Here, the $x$ direction is along the direction of the grating vector, and the $y$ direction is along the direction of the grating lines.

The phase, period, and tilt errors in different regions of the grating array will affect the diffraction wavefront, which will also be reflected in the interference fringes observed in the Fizeau interferometer. If there is a phase error, the Fizeau fringes will exhibit a displacement, and the diffraction wavefront will experience a discontinuity. If the diffraction wavefront discontinuity is $\delta$, the phase error $e_\text{p}$ (nm) is given by
\begin{equation}
	e_\text{p} = \frac{d_\text{g} \delta}{m\lambda_\text{F}}.
	\label{ep}
\end{equation}

If there is a tilt error, the Fizeau fringes will exhibit a change in fringe inclination, and the diffraction wavefront will experience a tilt relative to the reference wavefront. The normal vector of the reference wavefront $n_0$ is given by $(a_0, b_0, c_0)$ and the normal vector of the diffraction wavefront $n_m$ is represented as $(a_m, b_m, c_m)$. The tilt error $e_{\uptheta}$ (rad) can be expressed as:
\begin{equation}
	e_{\uptheta} = \arctan(\frac{d_\text{g}}{m\lambda_\text{F}}\tan(\gamma)),
	\label{et}
\end{equation}
where $\gamma$ is the angle between the vectors $n_0$ and $n_m$ in the $y$ direction.

In the presence of a period error, the Fizeau fringes will exhibit periodic variations, and the diffraction wavefront will experience a deflection. The period error $e_\text{d}$ (nm) is given by
\begin{equation}
	e_\text{d} = \frac{m\lambda_\text{F}}{\sin\left(\psi + \arcsin\left(\frac{m\lambda_\text{F}}{2d_\text{g}}\right)\right) + \frac{m\lambda_\text{F}}{2d_\text{g}}} - d_\text{g},
	\label{ed}
\end{equation}
where $\psi$ is the angle between the vectors $n_0$ and $n_m$ in the $x$ direction.

\subsection{Experimental results}
\begin{figure} [t]
	\begin{center}
		\includegraphics[height=9.7cm]{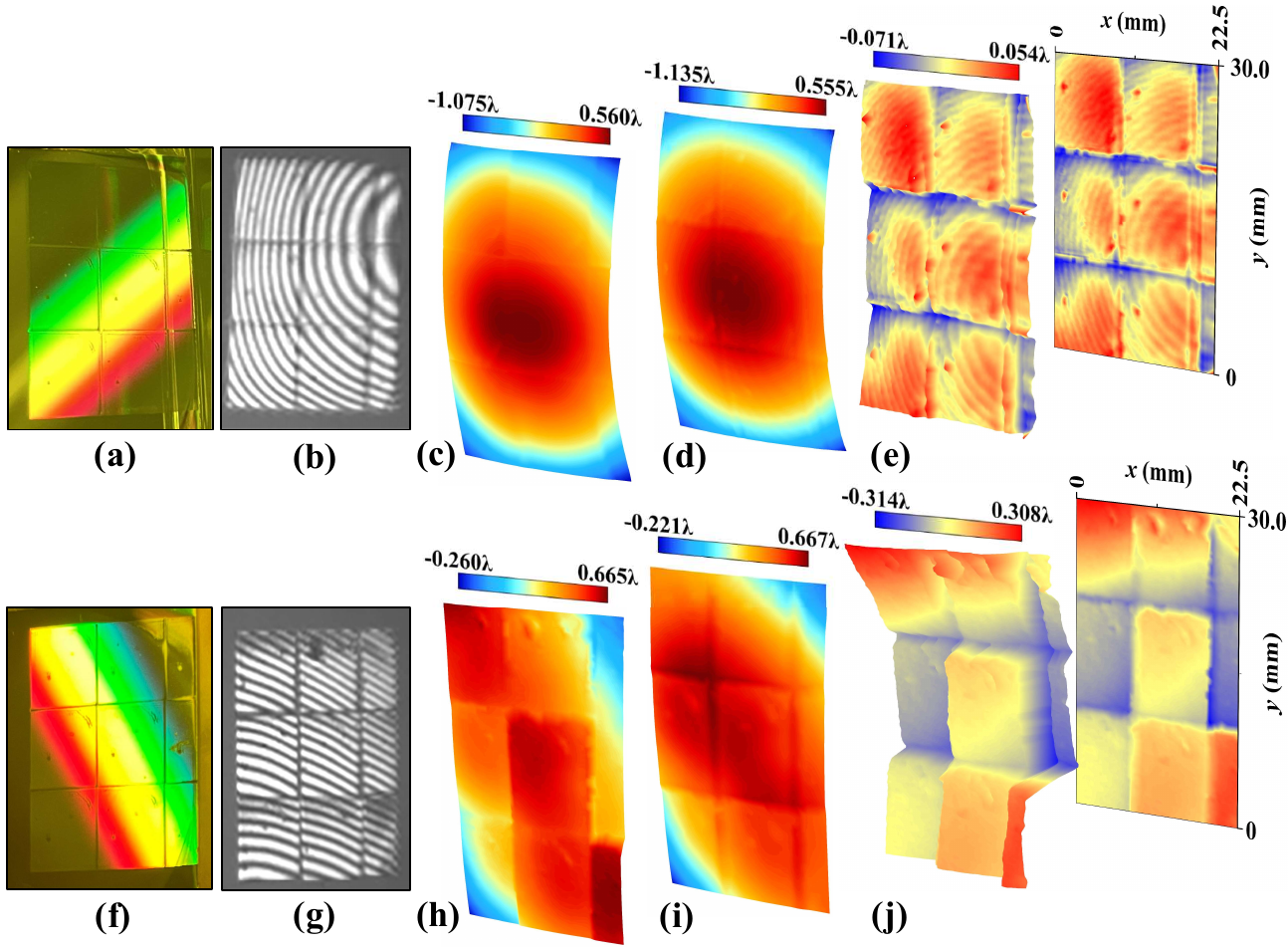}
	\end{center}
	\caption
	{ \label{result} Photograph, Fizeau fringes and wavefront measurement results for two samples: (a), (f) Photographs of Sample 1 and 2; (b), (g) Fizeau fringes of Sample 1 and 2; (c), (h) $-1$st-order diffraction wavefronts of Sample 1 and 2; (d), (i) Non-flatness of the substrates of Sample 1 and 2; (e), (j) $-1$st-order diffraction wavefronts of Sample 1 and 2 with substrate flatness error removed. (Sample 1: PV = 0.125 $\uplambda$, RMS = 0.023 $\uplambda$; Sample 2: PV = 0.621 $\uplambda$, RMS = 0.105 $\uplambda$).}
\end{figure}

Based on the above fabrication procedure, a $3 \times 3$ grating pattern array was fabricated on a photoresist-coated silicon substrate, with dimensions of (9+9+3.5) mm $\times$ (10+10+10) mm and a grating period of 1645 nm, as shown in Figure \ref{result}(a). 

The $-1$st-order diffraction wavefront was measured using a Fizeau interferometer with an operating wavelength of $\uplambda = 632.8$ nm. The Fizeau fringes exhibit good continuity, as shown in Figure \ref{result}(b), reflecting the uniformity of the diffraction wavefront. Figure \ref{result}(c) shows the results of the direct measurement of the $-1$st-order diffraction wavefront. The substrate used here is a silicon wafer, which is relatively thin and exhibits noticeable bending, as shown in Figure \ref{result}(d). We calculated the $-1$st-order diffraction wavefront after removing the substrate flatness error (Figure \ref{result}(e)). After eliminating the influence of the substrate, the peak-valley (PV) and root mean square (RMS) values of the grating array's $-1$st-order diffraction wavefront are 0.125 $\uplambda$ and 0.023 $\uplambda$, respectively. The wavefront measurement results indicate that the fabricated grating array is of high quality. 

The other $3 \times 3$ grating pattern array was fabricated without using our alignment system (Sample 2), as shown in Figure \ref{result}(f)-(j). It can be observed that the Fizeau fringes show noticeable discontinuities, and the $-1$st-order diffraction wavefront exhibits considerable discontinuities and deflections. The PV and RMS values of the second grating array's $-1$st-order diffraction wavefront with substrate flatness error removed are 0.621 $\uplambda$ and 0.105 $\uplambda$, respectively. 
\begin{figure} [htbp]
	\begin{center}
		\includegraphics[height=4.6cm]{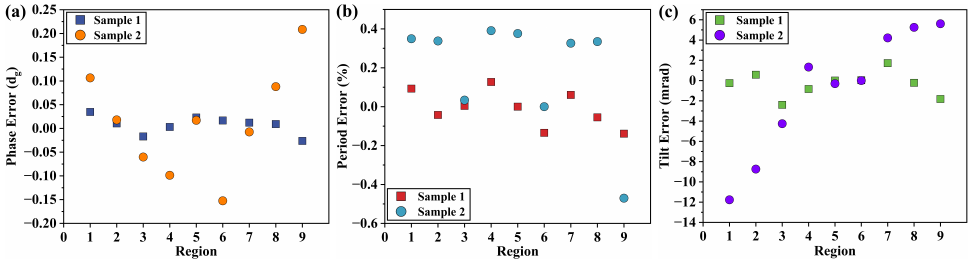}
	\end{center}
	\caption{\label{errors} Alignment errors relative to the reference flat for Sample 1 and 2: (a) Phase errors; (b) Period errors; (c) Tilt errors.}
\end{figure} 

The alignment errors for Sample 1 and 2 are also analyzed here. Utilizing Equations \ref{ep}, \ref{et}, and \ref{ed}, we calculate the phase, period, and tilt errors for each region relative to a reference flat, as presented in Figure \ref{errors}. The period, tilt errors and some random fabrication defects also affect the phase error. Therefore, the center of each region is selected to illustrate the phase error by calculating the mean value of the wavefront displacement. In the grating arrays, the maximum average phase errors between any two regions are 0.061 $\mathrm{d_g}$ for Sample 1 and 0.361 $\mathrm{d_g}$ for Sample 2. The largest period errors between any two regions are 0.266 \% for Sample 1 and 0.861 \% for Sample 2. The greatest tilt errors between any two regions are 4.1 mrad for Sample 1 and 17.4 mrad for Sample 2.
\begin{figure} [htbp]
	\begin{center}
		\includegraphics[height=17cm]{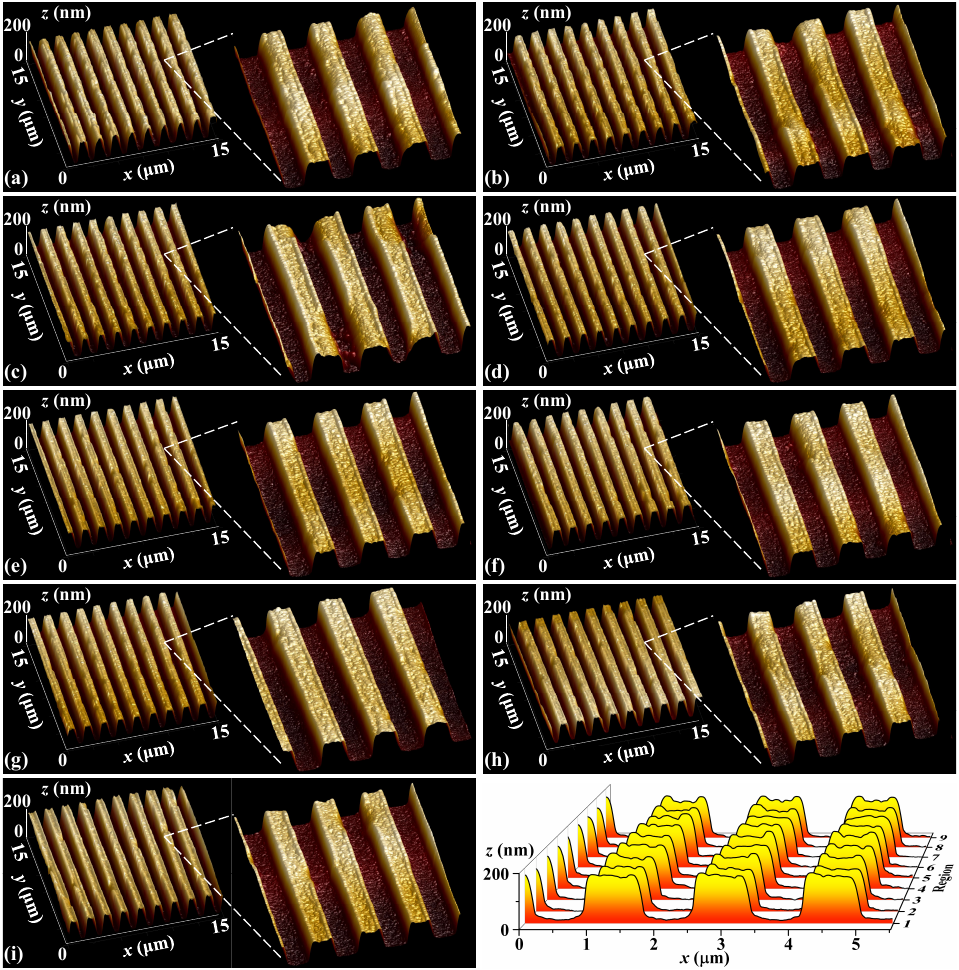}
	\end{center}
	\caption
	{\label{AFM} AFM images and cross-sectional views of the grating array: (a)-(i): AFM images of Regions 1-9; (j) Cross-sectional views of Regions 1-9.}
\end{figure}

We randomly select a measurement point in each region and utilize an atomic force microscope (AFM) to examine the microstructure of Sample 1, as illustrated in Figures \ref{AFM}(a)-(i). Figure \ref{AFM}(j) shows a cross-sectional view of the measurement points in each region. The AFM results demonstrate the consistency of the groove pattern across all areas.  

\begin{figure} [htbp]
	\begin{center}
		\includegraphics[height=12cm]{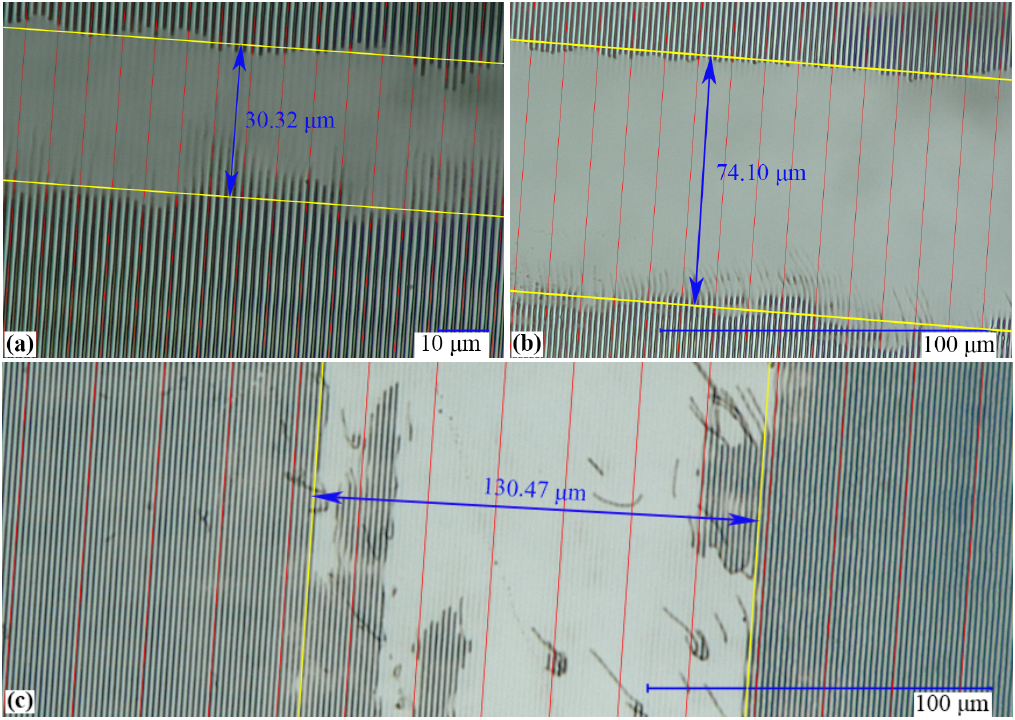}
	\end{center}
	\caption
	{Microscope images of the grating array seams, with the red lines representing reference lines with equal period and tilt angle: (a),(b) longitudinal seams;  (c) transverse seams.}
	\label{SEAM}
\end{figure}

Microscope images of some array seams are presented in Figure \ref{SEAM}. Figures \ref{SEAM}(a) and \ref{SEAM}(b) show the longitudinal seams, while Figure \ref{SEAM}(c) shows the transverse seams. The array seams between each region of the grating array are not larger than 236.25 \textmu m. These experimental results demonstrate the effectiveness of this method for fabricating large-area grating pattern arrays.

\section{Discussion}
We achieved global alignment reference using only reference gratings for fabricating large-area laser interference lithography pattern arrays. Two orthogonally placed reference gratings are used to monitor alignment errors. One of the reference gratings is separated from the substrate, allowing it to be reused for the subsequent rows of exposure after the substrate is moved longitudinally. In the lateral fabrication process, the exposure region is adjusted by moving the beam. For the vertical fabrication process, the substrate is fixed with a second reference grating, ensuring that the relative position and orientation between the substrate and exposure beam remains consistent across each row. This reference method can be applied in various types of laser interference lithography, aligning other forms of periodic pattern arrays.

In our experiment, a $3 \times 3$ grating pattern array of (9+9+3.5) mm $\times$ (10+10+10) mm was successfully fabricated with high quality, but the proposed method can be applied for the alignment of any number of rows and columns. This method provides an effective and convenient approach for fabricating large-area laser interference lithography pattern arrays.

\section{Methods}

\subsection{Exposure and Alignment System Parameters}
The laser source used in the system is a He-Cd laser with a wavelength of 441.6 nm and a power of 180 mW. Positive photoresist Shepliy S1805 was spin-coated onto the silicon wafers, resulting in a thickness of 200 nm. In the experiment, the exposure time was set to 18 s.

The apertures $\text{D}_1$ and $\text{D}_2$ were adjusted to create a beam area of 9 mm $\times$ 15 mm. Within this area, a 9 mm $\times$ 5 mm region is designated for monitoring the reference fringes generated by the reference grating $\mathrm{G_{R_1}}$, while a 5.5 mm $\times$ 5 mm region is used for monitoring the reference fringes generated from the reference grating $\mathrm{G_{R_2}}$. Consequently, the exposure area for Regions 1, 2, 4, 5, 6, and 7 is 9 mm $\times$ 10 mm, and 3.5 mm $\times$ 10 mm for Regions 3, 5, and 9. The reference grating $\mathrm{G_{R_1}}$ and $\mathrm{G_{R_2}}$ employed both have 600 lines/mm.

The CMOS sensor of the camera has a size of 11.26 mm × 5.98 mm and a resolution of 2048 px × 1088 px. The frame rate of the camera is 340 fps, while the acquisition rate in the experiment is 100 fps. The resolution of the piezoelectric actuator used for phase error compensation is 0.18 nm, with a range of 18 $\upmu$m. The angular resolution of the dual-axis piezoelectric mirror mount used for tilt and period error compensation is 0.7 $\upmu$rad, with a range of $\pm$ $3.5^\circ$. The positioning accuracy of the displacement stages $\text{S}_1$ and $\text{S}_2$ in the system is 25 $\upmu$m, with a repeatability of 3 $\upmu$m and a travel range of 300 mm.

\bibliography{report} 

\begin{thebibliography}{10}

\bibitem{cotA}
Fruncillo, S., Toh, Y.~T., Blanford, C.~F., Su, X., Liu, H., and Wong, L.~S.,
  ``Lithographic patterning of nanoscale arrays of the oxidase enzyme cota:
  Effects on activity and stability,'' {\em Advanced Materials
  Technologies}~{\bf 7}(8),  2200490 (2022).

\bibitem{ma2024nanofabrication}
Ma, R., Zhang, X., Sutherland, D., Bochenkov, V., and Deng, S.,
  ``Nanofabrication of nanostructure lattices: from high-quality large patterns
  to precise hybrid units,'' {\em International Journal of Extreme
  Manufacturing}~{\bf 6}(6),  062004 (2024).

\bibitem{barad2021large}
Barad, H.-N., Kwon, H., Alarc{\'o}n-Correa, M., and Fischer, P., ``Large area
  patterning of nanoparticles and nanostructures: current status and future
  prospects,'' {\em ACS nano}~{\bf 15}(4),  5861--5875 (2021).

\bibitem{ji2020patterning}
Ji, D., Li, T., and Fuchs, H., ``Patterning and applications of nanoporous
  structures in organic electronics,'' {\em Nano today}~{\bf 31},  100843
  (2020).

\bibitem{biochips}
Fruncillo, S., Su, X., Liu, H., and Wong, L.~S., ``Lithographic processes for
  the scalable fabrication of micro-and nanostructures for biochips and
  biosensors,'' {\em ACS sensors}~{\bf 6}(6),  2002--2024 (2021).

\bibitem{cheng2018high}
Cheng, P.-J., Huang, Z.-T., Li, J.-H., Chou, B.-T., Chou, Y.-H., Lo, W.-C.,
  Chen, K.-P., Lu, T.-C., and Lin, T.-R., ``High-performance plasmonic
  nanolasers with a nanotrench defect cavity for sensing applications,'' {\em
  Acs Photonics}~{\bf 5}(7),  2638--2644 (2018).

\bibitem{film}
Bagal, A., Zhang, X.~A., Shahrin, R., Dandley, E.~C., Zhao, J., Poblete, F.~R.,
  Oldham, C.~J., Zhu, Y., Parsons, G.~N., Bobko, C., et~al., ``Large-area
  nanolattice film with enhanced modulus, hardness, and energy dissipation,''
  {\em Scientific reports}~{\bf 7}(1),  9145 (2017).

\bibitem{2004spatial}
Moormann, C., Bolten, J., and Kurz, H., ``Spatial phase-locked combination
  lithography for photonic crystal devices,'' {\em Microelectronic
  engineering}~{\bf 73},  417--422 (2004).

\bibitem{wu2019large}
Wu, H., Jiao, Y., Zhang, C., Chen, C., Yang, L., Li, J., Ni, J., Zhang, Y., Li,
  C., Zhang, Y., et~al., ``Large area metal micro-/nano-groove arrays with both
  structural color and anisotropic wetting fabricated by one-step focused laser
  interference lithography,'' {\em Nanoscale}~{\bf 11}(11),  4803--4810 (2019).

\bibitem{shen2021}
Shen, H., Wang, Y., Cao, L., Xie, Y., Wang, Y., Zhang, Q., Zhang, W., Wang, S.,
  Han, Z., Zhu, X., et~al., ``Fabrication of periodical micro-stripe structure
  of polyimide by laser interference induced forward transfer technique,'' {\em
  Applied Surface Science}~{\bf 541},  148466 (2021).

\bibitem{huerta2017}
Huerta-Murillo, D., Aguilar-Morales, A.~I., Alamri, S., Cardoso, J.~T.,
  Jagdheesh, R., Lasagni, A.-F., and Oca{\~n}a, J.~L., ``Fabrication of
  multi-scale periodic surface structures on ti-6al-4v by direct laser writing
  and direct laser interference patterning for modified wettability
  applications,'' {\em Optics and Lasers in Engineering}~{\bf 98},  134--142
  (2017).

\bibitem{voisiat2019}
Voisiat, B., Wang, W., Holzhey, M., and Lasagni, A.~F., ``Improving the
  homogeneity of diffraction based colours by fabricating periodic patterns
  with gradient spatial period using direct laser interference patterning,''
  {\em Scientific reports}~{\bf 9}(1),  7801 (2019).

\bibitem{liu2023laser}
Liu, R., Cao, L., Liu, D., Wang, L., Saeed, S., and Wang, Z., ``Laser
  interference lithography—a method for the fabrication of controlled
  periodic structures,'' {\em Nanomaterials}~{\bf 13}(12),  1818 (2023).

\bibitem{zhou2016}
Zhou, Q., Li, X., Ni, K., Tian, R., and Pang, J., ``Holographic fabrication of
  large-constant concave gratings for wide-range flat-field spectrometers with
  the addition of a concave lens,'' {\em Optics Express}~{\bf 24}(2),  732--738
  (2016).

\bibitem{marczak2013direct}
Marczak, J., Rycyk, A., Sarzy{\'n}ski, A., Strzelec, M., Kusi{\'n}ski, J., and
  Major, R., ``Direct laser manufacturing of 1d and 2d micro-and submicro-scale
  periodic structures,'' in [{\em Laser Technology 2012: Applications of
  Lasers}{\nolinebreak\hspace{0.1em}]},   {\bf 8703},  103--114, SPIE (2013).

\bibitem{lasagni2024}
Lasagni, A. and Voisiat, B., ``The development of direct laser interference
  patterning: past, present, and new challenges,'' in [{\em Frontiers in
  Ultrafast Optics: Biomedical, Scientific, and Industrial Applications
  XXIV}{\nolinebreak\hspace{0.1em}]},   {\bf 12875},  68--76, SPIE (2024).

\bibitem{Telescope}
Barnes, S.~I., Cottrell, P.~L., Albrow, M.~D., Frost, N., Graham, G., Kershaw,
  G., Ritchie, R., Jones, D., Sharples, R., Bramall, D., Schmoll, J., Luke, P.,
  Clark, P., Tyas, L., Buckley, D. A.~H., and Brink, J., ``The optical design
  of the southern african large telescope high resolution spectrograph: Salt
  hrs,'' in [{\em Ground-based and Airborne Instrumentation for Astronomy
  II}{\nolinebreak\hspace{0.1em}]},  McLean, I.~S. and Casali, M.~M., eds.,
  {\em Proc. SPIE} {\bf 7014},  247--258, International Society for Optics and
  Photonics (2008).

\bibitem{steidel2022design}
Steidel, C., Peng, E., Fucik, J., Nash, R., Kaye, S., Jacoby, G., Delabre, B.,
  Sethuram, R., Divakar, D., Varshney, H.~M., et~al., ``Design and development
  of wfos, the wide-field optical spectrograph for the thirty meter
  telescope,'' in [{\em Ground-based and Airborne Instrumentation for Astronomy
  IX}{\nolinebreak\hspace{0.1em}]},   {\bf 12184},  706--719, SPIE (2022).

\bibitem{LaserFusion}
Clery, D., ``Laser fusion energy poised to ignite,'' {\em Science}~{\bf
  328}(5980),  808--809 (2010).

\bibitem{zuegel2006laser}
Zuegel, J., Borneis, S., Barty, C., Legarrec, B., Danson, C., Miyanaga, N.,
  Rambo, P., Leblanc, C., Kessler, T., Schmid, A., et~al., ``Laser challenges
  for fast ignition,'' {\em Fusion Science and Technology}~{\bf 49}(3),
  453--482 (2006).

\bibitem{xing2017spatially}
Xing, X., Chang, D., Hu, P., and Tan, J., ``Spatially separated heterodyne
  grating interferometer for eliminating periodic nonlinear errors,'' {\em
  Optics Express}~{\bf 25}(25),  31384--31393 (2017).

\bibitem{shimizu2021laser}
Shimizu, Y., ``Laser interference lithography for fabrication of planar scale
  gratings for optical metrology,'' {\em Nanomanufacturing and Metrology}~{\bf
  4}(1),  3--27 (2021).

\bibitem{Bonod:16}
Bonod, N. and Neauport, J., ``Diffraction gratings: from principles to
  applications in high-intensity lasers,'' {\em Adv. Opt. Photon.}~{\bf 8},
  156--199 (Mar 2016).

\bibitem{heilmann2001digital}
Heilmann, R.~K., Konkola, P.~T., Chen, C.~G., Pati, G., and Schattenburg,
  M.~L., ``Digital heterodyne interference fringe control system,'' {\em
  Journal of Vacuum Science \& Technology B: Microelectronics and Nanometer
  Structures Processing, Measurement, and Phenomena}~{\bf 19}(6),  2342--2346
  (2001).

\bibitem{SCAN}
Schattenburg, M.~L., Chen, C.~G., Heilmann, R.~K., Konkola, P.~T., and Pati,
  G.~S., ``{Progress toward a general grating patterning technology using
  phase-locked scanning beams},'' in [{\em Optical Spectroscopic Techniques,
  Remote Sensing, and Instrumentation for Atmospheric and Space Research
  IV}{\nolinebreak\hspace{0.1em}]},  Larar, A.~M. and Mlynczak, M.~G., eds.,
  {\em Proc. SPIE} {\bf 4485},  378 -- 384, International Society for Optics
  and Photonics (2002).

\bibitem{chen2002nanometer}
Chen, C.~G., Konkola, P.~T., Heilmann, R.~K., Joo, C., and Schattenburg, M.~L.,
  ``Nanometer-accurate grating fabrication with scanning beam interference
  lithography,'' in [{\em Nano- and Microtechnology: Materials, Processes,
  Packaging, and Systems}{\nolinebreak\hspace{0.1em}]},  Sood, D.~K., Malshe,
  A.~P., and Maeda, R., eds., {\em Proc. SPIE} {\bf 4936},  126--134 (2002).

\bibitem{TURUKHANO1996263}
Turukhano, B., Gorelik, V., Kovalenko, S., and Turukhano, N., ``Phase synthesis
  of a holographic metrological diffraction grating of unlimited length,'' {\em
  Optics $\&$ Laser Technology}~{\bf 28}(4),  263--268 (1996).

\bibitem{shi2010fabrication}
Shi, L. and Zeng, L., ``Fabrication of optical mosaic gratings by consecutive
  holographic exposures employing a latent-fringe based alignment technique,''
  in [{\em Holography, Diffractive Optics, and Applications
  IV}{\nolinebreak\hspace{0.1em}]},  {\em Proc. SPIE} {\bf 7848},  169--177
  (2010).

\bibitem{shi2011fabrication}
Shi, L. and Zeng, L., ``Fabrication of optical mosaic gratings: a
  self-referencing alignment method,'' {\em Optics express}~{\bf 19}(10),
  8985--8993 (2011).

\bibitem{ma2017achieving}
Ma, D., Zhao, Y., and Zeng, L., ``Achieving unlimited recording length in
  interference lithography via broad-beam scanning exposure with
  self-referencing alignment,'' {\em Scientific Reports}~{\bf 7}(1),  926
  (2017).

\end{thebibliography}
\bibliographystyle{spiebib}

\section{Acknowledgements}
This work was supported by the National Natural Science Foundation of China (No. 62275142) and Shenzhen Stable Supporting Program (No. WDZC20231124201906001).
\section{Author contributions}
X.G conceived the idea, conducted the experiments and testing, analyzed data, and wrote the manuscript. J.L and Z.J established the fringe locking system. X.L supported and supervised the work. All authors contributed to the manuscript preparation.
\section{Competing interests}
The authors declare no competing interests.

\end{document}